\documentclass[12pt]{article}
\usepackage{graphicx}
\usepackage{float}
\usepackage{epstopdf}
\begin{document}

\title{\bf Cosmological Analysis of Modified Holographic Ricci Dark Energy in Chern-Simons Modified Gravity}

\author{Sarfraz Ali \thanks{sarfraz270@ue.edu.pk}$~~$and M. Jamil Amir \thanks{mjamil.dgk@gmail.com}
\\Department of Mathematics, University of Education Lahore,\\Faisalabad Campus, Pakistan,\\
Department of Mathematics, Govt. Degree College Taunsa Sharif,\\ DG Khan, Pakistan.\\}

\date{}

\maketitle

\begin{abstract}
In this paper, we study the cosmological analysis of  the modified holographic Ricci dark energy model
and reconstruct different scalar field models in the context of Chern-Simon modified gravity. We
investigate the deceleration parameter, which shows that the universe is in the accelerating expansion
phase. The equation of state parameter in this case also favors the fact that dark energy is the dominant component
of universe which is responsible for the accelerated expansion. A number of scalar
fields, such as: quintessence, tachyon, K-essence and dilaton models are reconstructed using
modified holographic Ricci dark energy model in the context of dynamical CS modified gravity.
The quintessence and K-essence models represent exponentially increasing behaviors while Tachyon model
shows decreasing behavior. Unfortunately, the dilaton model has no numerical solution for modified
holographic Ricci dark energy model in the framework of dynamical Chern-Simon modified gravity.
\end{abstract}
~~~~PACS: 04.20.-q

\section{Introduction} A large number of evidences have been provided in the favor of the
accelerated expansion of the universe by Type Ia supernovae \cite{[1]}, Cosmic Microwave
Background (CMB) \cite{[2]}, weak lenseing \cite{[3]}, Large Scale Structures \cite{[4]}
and integrated Sachs-Wolfe effect \cite{[5]}. It is postulated that there exist a component
in the universe which has negative pressure is responsible for the accelerated expansion of
the universe, called Dark Energy (DE). The most familiar candidate of DE model is cosmological
constant $\Lambda$ which satisfy the cosmological observations \cite{[6]} but fails to resolve
the fine tuning problem and cosmic coincidence \cite{[7]}. In literature, there are many DE models
such as an evolving canonical scalar field \cite{[8]} quintessence, the phantom energy \cite{[9]},
quintom energy \cite{[10]} and so forth.

In recent studies to understand the nature of the universe, a new DE model has been constructed
in the context of quantum gravity on holographic principle named as holographic dark energy (HDE)
model \cite{[11]}. This principle is extensively used to study the quantum behavior of black holes.
The energy density of HDE is defined as $\rho=3c^{2}M^{2}_{pl} L^{-2}$, where $c$ is constant,
$M_{pl}$ plank mass and $L$ is supposed to be size of the universe. For the Hubble radius $H^{-1}$,
this HDE model density is very similar to the observational results.
Gao et al. \cite{[12]} motivated by the holographic principle, introduced a new DE model
which is inversely proportional to Ricci scalar curvature called Ricci Dark Energy (RDE). Their
investigation shows that this RDE model solve the causality problem and the evolution of density
perturbations of matter power spectra and CMB anisotropy is not much affected by such modification.
Granda and Oliveros \cite{[13]} introduced a new infrared cut-off for the HDE model and reconstruct
the potentials and fields for different DE models such as the quintessence,
tachyon, K-essence and dilaton for FRW universe. Karami and Fehri \cite{[14]} using Granda and Oliveros
cut-off studied the non-flat FRW universe to find the DE density, the deceleration parameter and the
equation of state (EoS).

Jackiw and Pi \cite{[15]} introduced Chern-Simons (CS) modified gravity in which the
Einstein-Hilbert action is modified as the sum of parity-violating CS term and scalar field.
Silva and Santos \cite{[16]} analysis the RDE of FRW universe and found it similar to GCG in
the context of CS modified gravity. Jamil and Sarfraz \cite{[17]} work the same for amended FRW universe
and  present their results graphically. Jawad and Sohail \cite{[18]} considering modified QCD ghost dark
energy model investigate the dynamics of scalar field and potentials of various
scalar field models in the framework of dynamical CS modified gravity. Jamil and Sarfraz
\cite{[19]} considering HDE model found the accelerated expansion behavior of the universe under
the certain restrictions on the parameter $\alpha$. We study the correspondence between the quintessence,
K-essence, tachyon and dilaton field models and holographic dark energy model.
Pasqua et al. \cite{[20]} investigated the HDE, modified holographic Ricci dark energy (MHRDE) and
another model which is a combination of higher order derivatives of the Hubble parameter in the framework of CS modified gravity.

In this paper, working on same lines using the MHRDE model, we explore the energy density, deceleration parameter,
EoS parameter and correspondence between different models. The paper organized in following order.
The basic formulism of CS modified gravity is discussed in section $2$. The section $3$, is devoted
for the investigation of energy density, deceleration parameter and EoS parameter. The correspondence
between scalar field models such that quintessence, tachyon, K-essence and dilaton model is given in
section $4$. Summery and concluding remarks are in the last section.

\section{ABC of Chern-Simon Modified Gravity}
Jackiw and Pi  modify the 4-dimensional GR theory introducing a Chern-Simon term in Einstein-Hilbert
action, is given by
\begin{eqnarray}
S=\int d^{4}x\sqrt{-g}[\chi R+\frac{\zeta}{4}\Theta~^{*}RR-\frac{\eta}{2}(g^{\mu\nu}\nabla_{\mu}\Theta\nabla_{\nu}\Theta+2V[\Theta])]+S_{mat},
\end{eqnarray}
the terms used in this relation are defined as $\chi = (16\pi G)^{-1}$, $R$ is the usual
Ricci scalar,  the term $^{*}RR$ is defined as $ ^{*}RR= {{^{*}R^a}_b}^{cd} {R^b}_{acd}$,
is topological invariant, called Pontryagin term where ${{^{*}R^a}_b}^{cd}$ is the dual
of Reimann tensor ${R^b}_{acd}$ can be expressed as
${{^{*}R^a}_b}^{cd}=\frac{1}{2}\epsilon^{cdef}{R^a}_{bef}$, the $\nabla_{\mu}$ is titled as
covariant derivative, the dimensionless parameters $\zeta$ and $\eta$ are treated here as
coupling constants and $V[\Theta]$ is potential and in the context of string theory it is assumes that
$V[\Theta]=0$ . The most important term is $\Theta$, called CS coupling field works as
a deformation function of the spacetime which is always other than constant (If $\Theta$
is constant function then the CS theory reduces to GR).

The variation of Einstein-Hilbert action $S$ corresponding to metric tensor $g_{\mu\nu}$ and scalar field $\Theta$
resulted into a set of field equations of CS modified gravity given by
\begin{eqnarray}
G_{\mu\nu}+l C_{\mu\nu} &=& \chi T_{\mu\nu},\\
g^{\mu \nu}\nabla_{\mu}\nabla_{\nu}\Theta &=&-\frac{\zeta}{4}~^{*}RR,
\end{eqnarray}
where $G_{\mu\nu}$, $C_{\mu\nu}$, $l$, and $T_{\mu\nu}$ are called Einstein tensor, Cotton tensor
(C-tensor), coupling constant and  energy-momentum tensor respectively. The energy-momentum tensor
is a combination of matter part $T^{m}_{\mu\nu}$ and the external field part $T^{\Theta}_{\mu\nu}$.
The mathematical expressions for these terms are followed as
\begin{eqnarray}
C^{\mu\nu}&=&-\frac{1}{2\sqrt{-g}}[\upsilon_{\sigma}\epsilon^{\sigma\mu\zeta\eta}\nabla_{\zeta}R^{\nu}_{\eta}+
\frac{1}{2}\upsilon_{\sigma\tau}\epsilon^{\sigma\nu\zeta\eta}R^{\tau\mu}_{\zeta\eta}]+(\mu\longleftrightarrow\nu),
\\
T^{m}_{\mu\nu}&=&(\rho+p)u_\mu u_\nu-p g_{\mu\nu},\\
T^{\Theta}_{\mu\nu}&=&\zeta(\partial_{\mu}\Theta)(\partial_{\nu}\Theta)-\frac{\zeta}{2}g_{\mu\nu}(\partial^\lambda\Theta)(\partial_{\lambda}\Theta).
\end{eqnarray}
Here $\rho$ is energy density, $p$ is pressure and $u_{\mu} = (1,0,0,0)$ denote standard
time-like 4-velocity respectively. The terms $\upsilon_{\sigma}\equiv\nabla_{\sigma}\Theta$, $\upsilon_{\sigma\tau}\equiv\nabla_{\sigma}\nabla_{\tau}\Theta$.
On the basis of choice of $(\zeta\neq0$ and $\chi\neq0)$ and $\chi\neq0$ and $\zeta=0)$,
this theory is divided into two distinct theories named as dynamical and non-dynamical
Chern-Simon Modified gravity respectively.

\section{Modified Holographic Ricci Dark Energy Model}
In this paper we study the FRW universe in the framework of dynamical CS modified gravity.
The 00-component of field equation of FRW universe using Eq.(2), we get,
\begin{eqnarray}
H^{2}=\frac{1}{3}\rho_{D}+\frac{1}{6}\dot{\Theta}^{2}.
\end{eqnarray}
Here $H=\frac{\dot{a}}{a}$ is called Hubble parameter and $\dot{a}$ is time derivative of scale factor $a(t)$.
The Pontryagin term $^{*}RR$  vanishes for FRW metric identically, so, Eq.(3) takes the form
\begin{eqnarray}
g^{\mu\nu}\nabla_{\mu}\nabla_{\nu}\Theta=g^{\mu\nu}[\partial_{\mu}\partial_{\nu}\Theta-\Gamma^{\rho}_{\mu\nu}\partial_{\rho}\Theta]=0.
\end{eqnarray}
Keeping in view, the $\Theta$ is a function of spacetime, we consider, $\Theta=\Theta(t)$, Eq.(8) turns out to be.
\begin{eqnarray}
\dot{\Theta}=C a^{-3}.
\end{eqnarray}
$C$ is constant of integration other than zero (as if this $C$ is zero then the function $\Theta$ become
constant which reduces the CS theory to GR).

Using holographic principle, Hooft \cite{[22]} proposed very simple and convenient model
to investigate the issues raised in DE, named as HDE model. This model is used in different
scenarios such that Hubble radius and cosmological conformal time of particle horizon \cite{[23],[27]}.
An interesting holographic RDE model defined as $L= |R|^{-\frac{1}{2}}$, where $R$ is Ricci
curvature scalar, was purposed by Gao et al \cite{[28]}.

In this paper, we use HDE model suggested by Granda and Oliveros in \cite{[29]} defined as
\begin{eqnarray}
\rho_{MHRDE}=\frac{2}{\alpha-\beta}(\dot{H}+\frac{3\alpha}{2}H^{2}),
\end{eqnarray}
here $\alpha$ and $\beta$ are constants and in limiting case if $(\alpha=\frac{4}{3},\beta=1)$, the
Granda and Oliveros IR cut-off reduces to Gao et.al. IR cut-off.
Using Eq.(9) and Eq.(10) in Eq.(7) we arrive at
\begin{eqnarray}
H^{2}=\frac{2}{3(\alpha-\beta)}(\dot{H}+\frac{3\alpha}{2}H^{2})+\frac{1}{6}C^{2} a^{-6}
\end{eqnarray}
and solving the differential equation, the scale factor $a(t)$ is explored as
\begin{eqnarray}
a(t) =[\frac{12C_{1}}{C(\alpha-4)}-\frac{3C(\alpha-4)}{4}t^{2}]^\frac{1}{6}.
\end{eqnarray}
To avoid the singular solution, it is provided that $\alpha\neq4$.
The Hubble parameter $H(t)=\frac{\dot{a}}{a}$ can be evaluated as
\begin{eqnarray}
H(t)=-\frac{Ct(\alpha-4)}{4\Big(\frac{12C_{1}}{C(\alpha-4)}-\frac{3}{4}Ct^{2}(\alpha-4)\Big)}.
\end{eqnarray}
Since, the scale factor has been explored by assuming  $\beta=4$ in Eq.(10) so it takes the form
\begin{eqnarray}
\rho_{MHRDE}=\frac{2}{\alpha-4}(\dot{H}+\frac{3\alpha}{2}H^{2}),
\end{eqnarray}
Using corresponding values of $H(t)$ and $\dot{H}(t)$, the expression for MHRDE density turned out to be
\begin{eqnarray}
\rho_{MHRDE}=\frac{(\alpha-4)\Big(C^{4}t^{2}(\alpha-4)^{2}(\alpha-2)-32C^{2}C_{1}\Big)}{3\Big(C^{2}t^{2}(\alpha-4)^{2}-16C_{1}\Big)^{2}}.
\end{eqnarray}
In term of redshift parameter the density is given as
\begin{eqnarray}
\rho(z)=\frac{1}{4}(1 + z)^{6} [-C(\alpha-2) + 12 C_{1} (1 + z)^{6}].
\end{eqnarray}
The energy density of this model is increasing for all values of  $C$, $\alpha<2$ and $C_{1}>0$.
We plot a graph for different values of these parameters. 
\begin{figure}[H]
\centering
\includegraphics[width=3.5in]{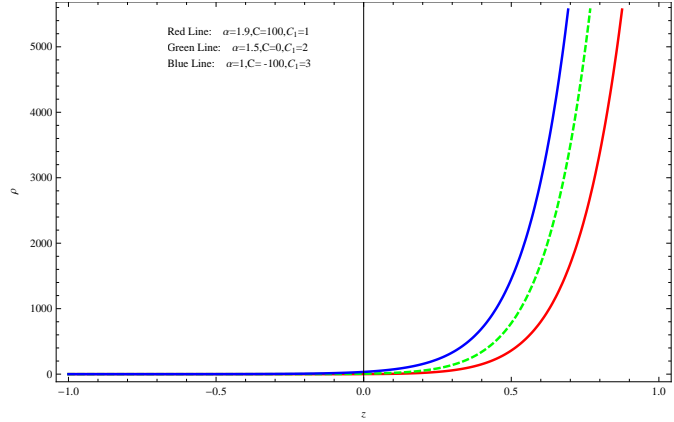}
\caption{\small{Density vs Redshift Parameter}}\label{pic01}.
\end{figure}
The graphical behavior of the density is exponentially increasing after $z>0$ in the context of
CS gravity using this MHRDE model.
\subsection{Deceleration Parameter}
The rate of expansion of the universe remained unchanged at constant values of  $\dot{a}(t)$ 
and deceleration term $q$ along with condition imposed on scale factor $a(t)$ that is
$a(t)\propto t$, where $t$ is cosmic time. The Hubble parameter $H$ remain constant and
deceleration term $q=-1$, when de Sitter and steady-state universes are  under consideration.
Furthermore, the deceleration parameter varies with time for some universes available in literature.
Using the variational values of $H$ and $q$,  we can classify all the defined universe models whether,
they are in expansion or contraction mode, acceleration or deceleration mode:\\
\textbf{1}.~~$H>0,~~~q>0$,~~~~~~expanding and decelerating\\
\textbf{2}.~~$H>0,~~~q=0$,~~~~~~expanding, zero deceleration\\
\textbf{3}.~~$H<0,~~~q=0$,~~~~~~contracting, zero deceleration\\
\textbf{4}.~~$H<0,~~~q>0$,~~~~~~contracting and decelerating\\
\textbf{5}.~~$H<0,~~~q<0$,~~~~~~contracting and accelerating\\
\textbf{6}.~~$H=0,~~~q=0$,~~~~~~static.\\
The deceleration parameter $q$ in term of Hubble parameter $H$ is defined as
\begin{eqnarray}
q(t)=-1-\frac{\dot{H}}{H^2},
\end{eqnarray}
where $\dot{H}$ represent the derivative of $H$ with respect to t, termed as
\begin{eqnarray}
\dot{H}(t)=-\frac{C^{2}(\alpha-4)^{2}\Big(C^{2}t^{2}(\alpha-4)^{2}+16C_{1}\Big)}{3\Big(C^{2}t^{2}(\alpha-4)^{2}-16C_{1}\Big)^{2}}.
\end{eqnarray}
Substituting these values in Eq.(17), we find deceleration parameter $q$,
\begin{eqnarray}
q(t)=2+\frac{48 C_{1}}{(\alpha-4)^{2}C^{2}t^2}.
\end{eqnarray}
Representing in the form of redshift, it  becomes
\begin{eqnarray}
q(z)=\frac{2[C (\alpha-4) - 30 C_{1}(1 + z)^{6}]}{C(\alpha-4) - 12C_{1}(1 + z)^{6}}
\end{eqnarray}
Its obvious that the deceleration parameter depends on $C, C_{1}, \alpha$ and redshift parameter $z$.
At present epoch $z=0$, the deceleration parameter $q<0$ for each of case given below
\begin{eqnarray}
\textbf{1}.~~~\alpha<4,~~~C_{1}<0~~~ \textbf{and}~~~\frac{12C_{1}}{\alpha-4}<C<\frac{30C_{1}}{\alpha-4},\nonumber\\
\textbf{2}.~~~\alpha>4,~~~C_{1}<0~~~ \textbf{and}~~~\frac{30C_{1}}{\alpha-4}<C<\frac{12C_{1}}{\alpha-4},\nonumber\\
\textbf{3}.~~~\alpha<4,~~~C_{1}>0~~~ \textbf{and}~~~\frac{30C_{1}}{\alpha-4}<C<\frac{12C_{1}}{\alpha-4},\nonumber\\
\textbf{4}.~~~\alpha>4,~~~C_{1}>0~~~ \textbf{and}~~~\frac{12C_{1}}{\alpha-4}<C<\frac{30C_{1}}{\alpha-4}.
\end{eqnarray}
For all these conditions, the modal under consideration in CS gravity advocates that the
universe is in accelerated expansion phase.
\subsection{Equation of State Parameter}
The nature of component which is dominating univirse can be study with the EoS parameter $\omega$.
In fact, it illustrate the era of dominance of universe by certain component.
For example, $\omega=0,\frac{1}{3}$ and $1$ predict that the universe
is under dust, radiation and stiff fluid influence respectively. While $\omega=-\frac{1}{3},-1$
and $\omega<-1$ stand for quintessence DE, $\Lambda$CDM and Phantum eras respectively.
Now, differentiating Eq.(15) with respect to time $t$
\begin{eqnarray}
\dot{\rho}(t)=-\frac{2C^{4}t(\alpha-4)^{3}\Big(C^{2}t^{2}(\alpha-4)^{2}(\alpha-2)+16C_{1}(\alpha-6)\Big)}{3\Big(C^{2}t^{2}(\alpha-4)^{2}-16C_{1}\Big)^{3}}.
\end{eqnarray}
The conservation equation in CS modified gravity in given by \cite{[31]}
\begin{eqnarray}
\dot{\rho}+3 H(\rho + p)=0.
\end{eqnarray}
The expression for the EoS parameter $\omega$ can be explore using Eq.(23) such that
\begin{eqnarray}
\omega(t)=-1-\frac{\dot{\rho}(t)}{3H(t)\rho(t)}.
\end{eqnarray}
Making use of Eq.(13), Eq.(18) and Eq.(22) in Eq.(24), we found analytic solution of EoS parameter as given below
\begin{eqnarray}
\omega(t)=-1-\frac{-2C^{2}t^{2}(\alpha-4)^{2}(\alpha-2)-32C_{1}(\alpha-6)}{C^{2}t^{2}(\alpha-4)^{2}(\alpha-2)-32C_{1}}.
\end{eqnarray}
The EoS parameter $\omega$ in term of redshift parameter $z$ look like
\begin{eqnarray}
 \omega(z)= \frac{C(\alpha-2)-36C_{1}(1 + z)^{6}}{C(\alpha-2)-12C_{1}(1 + z)^{6}}.
\end{eqnarray}
Obviously, EoS parameter $\omega$ is a function of variable redshift parameter $z$ depends on $C$, $C_{1}$ and $\alpha$.
At the present epoch $z=0$ the EoS is $\omega<-1$ for each of the case described below
\begin{eqnarray}
\textbf{1}.~~~C < 0, ~~~ C_{1} < 0~~~ \textbf{and} ~~~ \frac{12C_{1} + 2C}{C} < \alpha < \frac{24C_{1}+2C}{C},\nonumber\\
\textbf{2}.~~~C < 0, ~~~ C_{1} > 0~~~ \textbf{and} ~~~ \frac{24C_{1}+2C}{C} < \alpha < \frac{12C_{1} + 2C}{C},\nonumber\\
\textbf{3}.~~~C > 0, ~~~ C_{1} < 0~~~ \textbf{and} ~~~ \frac{24C_{1}+2C}{C} < \alpha < \frac{12C_{1} + 2C}{C},\nonumber\\
\textbf{4}.~~~C > 0, ~~~ C_{1} > 0~~~ \textbf{and} ~~~ \frac{12C_{1} + 2C}{C} < \alpha < \frac{24C_{1}+2C}{C}.
\end{eqnarray}
For different values of these parameters, the EoS $\omega<-1$ favor the fact that universe is dominated by DE.
\section{Study of MHRDE Model Using Scaler Field Models}
In this section, we discuss different scalar field models like quintessence, tachyon,
K-essence and dilaton models in the framework of CS modified gravity.
To study, the behavior of quantum gravity, we explore the potential and scalar field.
\subsection{Quintessence Model}
A DE model is developed to explain the late-time cosmic acceleration called quintessence,
is a simplest scalar field which have no theoretical problem like ghosts and Laplacian instabilities
appearance \cite{[33]}. This model is useful to settle down the issue of fine tuning in cosmology
considering the time dependent EoS. Using this model, we can explain the cosmic acceleration having
negative pressure when potential energy dominates the kinetic energy. The energy and pressure densities are defined as
\begin{eqnarray}
\rho_{Q}=\frac{1}{2}\dot{\phi}^{2}+V(\phi), ~~~~~~~~ p_{Q}= \frac{1}{2}\dot{\phi}^{2}-V(\phi),
\end{eqnarray}
where the scalar field $\phi$ is differentiated with respect to $t$.
The EoS parameter for the quintessence is
\begin{eqnarray}
\omega_{\phi}=\frac{\frac{1}{2}\dot{\phi}^{2}-V(\phi)}{\frac{1}{2}\dot{\phi}^{2}+V(\phi)}.
\end{eqnarray}

The comparison of EoS $\omega_{\phi}$ formulated for quintessence model given in Eq.(29) and
EoS $\omega$ calculated for MHRDE modal given in Eq.(25)turned as
\begin{eqnarray}
\frac{\frac{1}{2}\dot{\phi}^{2}-V(\phi)}{\frac{1}{2}\dot{\phi}^{2}+V(\phi)}= \frac{C^{2}t^{2}(\alpha-4)^{2}(\alpha-2)+32C_{1}(\alpha-5)}{C^{2}t^{2}(\alpha-4)^{2}(\alpha-2)-32C_{1}}.
\end{eqnarray}
Now, equating density of quintessence model given in Eq.(28) and density evaluated from MHRDE
model represented in Eq.(15) expressed as
\begin{eqnarray}
\frac{1}{2}\dot{\phi}^{2}+V(\phi)=\frac{(\alpha-4)\Big(C^{4}t^{2}(\alpha-4)^{2}(\alpha-2)-32C^{2}C_{1}\Big)}{3\Big(C^{2}t^{2}(\alpha-4)^{2}-16C_{1}\Big)^{2}}.
\end{eqnarray}
Using Eq.(30) and (31), we arrive at
\begin{eqnarray}
\dot{\phi}^{2}=\frac{2C^{2}(\alpha-4)\Big(C^{2}t^{2}(\alpha-4)^{2}(\alpha-2)+16C_{1}(\alpha-6)\Big)}{3\Big(C^{2}t^{2}(\alpha-4)^{2}-16C_{1}\Big)^{2}}
\end{eqnarray}
and integrating the last equation with respect to $t$
\begin{eqnarray}
\phi(t)&=&\sqrt{\frac{2}{3}}\Bigg[-\sqrt{2}\tanh^{-1}\Big(\frac{\sqrt{2}Ct(\alpha-4)^{\frac{3}{2}}}{\sqrt{C^{2}t^{2}(\alpha-4)^{2}(\alpha-2)+16C_{1}(\alpha-6)}}\Big)\nonumber\\
&+&\sqrt{\frac{\alpha-2}{\alpha-4}}\ln\Bigg[C\Big(Ct(\alpha^{2}-2\alpha+8)+\sqrt{\alpha-2}\nonumber\\
&\times&\sqrt{C^{2}t^{2}(\alpha-4)^{2}(\alpha-2)+16C_{1}(\alpha-6)}\Big)\Bigg]\Bigg].
\end{eqnarray}
In terms of redshift parameter $z$, it turns out to be
\begin{tiny}
\begin{eqnarray}
\phi(z)&=&-\frac{2}{3(\alpha-4)}\tanh^{-1}\Bigg[\frac{(\alpha-4)^{\frac{3}{2}}\sqrt{\frac{4C}{(1+z)^{6}}+\frac{48C_{1}}{4-\alpha}}}{\sqrt{8-2\alpha}\sqrt{\frac{(\alpha-4)(-c(\alpha-2)+24(1 + z)^{6}C_{1})}{(1 + z)^{6}}}}\Bigg]\nonumber\\
&+&(\alpha-4)\sqrt{\frac{2(\alpha-2)}{3}}\log\Bigg[\frac{2C}{\sqrt{3}}\Big(\sqrt{\alpha-2}\sqrt{\frac{(\alpha-4)(-c(\alpha-2)+24(1+z)^{6}C_{1})}{(1+z)^{6}}}\nonumber\\
&+&\frac{\sqrt{c}(8-6\alpha+\alpha^{2})\sqrt{\frac{4}{(1+z)^{6}}+\frac{48C_{1}}{C(4-\alpha)}}}{2\sqrt{4-\alpha}}\Big)\Bigg].
\end{eqnarray}
\end{tiny}
Again using Eq.(30) and Eq.(31), the potential for quintessence model can be explored as
\begin{eqnarray}
V(t)=-\frac{16C^{2}C_{1}(\alpha-4)^{2}}{3(C^{2}t^{2}(\alpha-4)^{2}-16C_{1})^{2}}.
\end{eqnarray}
and making it convenient to discuss we change into redshift parameter
\begin{eqnarray}
V(z)=-3C_{1}(1+z)^{12}.
\end{eqnarray}
It is obvious that the potential of quintessence model depends only on the values of constant
of integration $C_{1}$. It shows the increasing behavior for all $C_{1}<0$ and decreasing
for $C_{1}>0$. It is interesting that the parameters $C$ and $\alpha$ are not appeared in
the final expression of potential in the frame work of CS Modified gravity.
\subsection{Tachyon Model}
Much attention have been given to tachyon field models in last few decades in string theory and cosmology \cite{[34]}-\cite{[40]}.
In fact, isotropic cosmological models whose radius depends on time and their
potential can be constructed using minimally coupled scalar field model \cite{[41]}.
Since the same procedure for the correspondence between minimally coupled scalar field models and tachyon can be utilize
to study the similar cosmological evolution \cite{[40]}.
The energy and pressure densities for the tachyone fiels model are expressed as
\begin{eqnarray}
\rho=\frac{V(\phi)}{\sqrt{1-\dot{\phi}^{2}}},~~~~~~ p=-V(\phi)\sqrt{1-\dot{\phi}^{2}}.
\end{eqnarray}
Since $p=\rho \omega$, using the above expressions, the EoS parameter can be evaluated as
\begin{eqnarray}
\omega=\dot{\phi}^{2}-1
\end{eqnarray}
The comparison of Eq.(25) with Eq.(38) gives  kinetic energy $\phi(t)$  such that
\begin{eqnarray}
\phi(t)= \sqrt{\frac{C_{1}(6-\alpha)}{C^{2}(\alpha-4)^{2}(\alpha-2)}} E[\arcsin(t\sqrt{\frac{C^{2}(\alpha-4)^{2}(\alpha-2)}{32C_{1}}});\frac{2}{\alpha-6}].
\end{eqnarray}
The tachyon potential in this case is
\begin{tiny}
\begin{eqnarray}
V(t)=\frac{C^{2}(\alpha-4)}{3(C^{2}t^{2}(\alpha-4)^{2}-16C_{1})^{2}}\times \sqrt{[C^{2}t^{2}(\alpha-4)^{2}(\alpha-2)-32C_{1}][C^{2}t^{2}(\alpha-4)^{2}(\alpha-2)+32C_{1}(\alpha-5)]}.
\end{eqnarray}
\end{tiny}
Now, the kinetic and potential energies of tachyon model in term of redshift parameter $z$ respectively,
\begin{eqnarray}
\phi(z)&=& \frac{4\sqrt{2C_{1}(\alpha-6)}}{\sqrt{C^{2}(\alpha-2)(\alpha-4)^{2}}}EllipticE\Bigg[\sin^{-1}\Big[\nonumber\\
&&\frac{\sqrt{\frac{C^{2}(\alpha-2)(\alpha-4)^{2}}{C_{1}}}\sqrt{\frac{4}{(1+z)^{6}}+\frac{48C_{1}}{C(4-\alpha)}}}{4\sqrt{6C(4-\alpha)}}\Big],-\frac{2}{\alpha-6}\Bigg]
\end{eqnarray}
and
\begin{eqnarray}
V(z)&=&-\frac{1}{4}(1 + z)^{6}\Big(C(\alpha-2)-12(1 + z)^{6}C_{1}\Big)\nonumber\\
&&\sqrt{1-\frac{1}{2}(1 + z)^{6}\Big(-c(\alpha-2)+24(1+z)^{6}C_{2})}.
\end{eqnarray}
To investigate the behavior of potential ,we plot a graph of $V(z)$ vs $z$ 
\begin{figure}[H]
\centering
\includegraphics[width=4in]{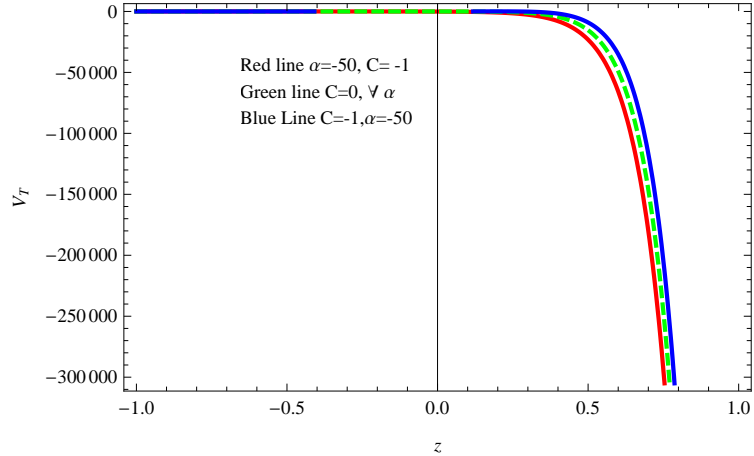}
\caption{\small{Potential vs Redshift Parameter}}\label{pic02}.
\end{figure}
The graph plotted for the potential $V(z)$ of tachyon model against redshift $z$ parameter
shows the decreasing behavior irrespective of the values of parameters $\alpha$, $C$ and $C_{1}$.
This graph is plotted by taking particular values of these parameters to elaborate the result. 
\subsection{K-essence}
Armendariz et al. \cite{[21]} introduced the dynamical concept of “k-essence” to explain the fact
of accelerated expansion of universe. This model solve the fine tuning problem of parameters.
Actually, k-essence is developed on the principle of dynamical attractor solution thats why it works
as cosmological constant at the onset of matter domination. The energy and pressure densities of this model are given as
\begin{eqnarray}
\rho=V(\phi)(-X+3X^{2}),~~~~~~~~~p=V(\phi)(-X+X^{2}).
\end{eqnarray}
Where $X=\frac{\dot{\phi}^{2}}{2}$, the EoS parameter for this model is
\begin{eqnarray}
\omega_=\frac{1-X}{1-3X}
\end{eqnarray}
Equating Eq.(25) and Eq.(44), we obtain
\begin{eqnarray}
X(t)=\frac{16C_{1}(\alpha-4)}{C^{2}t^{2}(\alpha-4)^{2}(\alpha-2)+16C_{1}(3\alpha-14)}.
\end{eqnarray}
Since $X(t)=\frac{\dot{\phi}^{2}}{2}$, the integration of the above equation provides $\phi(t)$
\begin{eqnarray}
\phi(t)&=&\frac{4C_{1}}{\sqrt{(\alpha-4)(\alpha-2)}}\ln[C\Big(Ct(\alpha^{2}-6\alpha+8)\nonumber\\
&+&\sqrt{\alpha-2}\sqrt{C^{2}t^{2}(\alpha-4)^{2}(\alpha-2)+16C_{1}(3\alpha-14)}\Big)].
\end{eqnarray}
The kinetic energy $\phi$ in term of redshift parameter $z$ is turned as
\begin{eqnarray}
\phi(z)&=&\frac{4C_{1}}{c\sqrt{\alpha-2}\sqrt{(\alpha-4)C_{1}}}\nonumber\\
&\times&\log\Bigg[\frac{2C}{\sqrt{3}}\sqrt{\frac{(\alpha-4)\Big(-C(\alpha-2)+48(1+z)^{6}C_{1}\Big)}{(1+z)^{6}}}\nonumber\\
&+&\frac{\sqrt{C}(8-6\alpha+\alpha^{2})\sqrt{\frac{4}{(1+z)^{6}}+\frac{48C_{1}}{C(4-\alpha)}}}{2\sqrt{4-\alpha}}\Bigg].
\end{eqnarray}
The k-essence potential is calculated using Eq.(43),(44) and Eq.(25) as
\begin{eqnarray}
V(t)=-\frac{C^{2}\Big(C^{2}t^{2}(\alpha-4)^{2}(\alpha-2)+16C_{1}(3\alpha-14)\Big)^{2}}{48C_{1}\Big(C^{2}t^{2}(\alpha-4)^{2}-16C_{1}\Big)^{2}}.
\end{eqnarray}
Now, we convert this function in term of redshift parameter $z$ and investigate its behavior.
\begin{eqnarray}
V(z)=-\frac{\Big(c(\alpha-2)-48(1 + z)^{6}C_{1}\Big)^{2}}{48C_{2}}.
\end{eqnarray}
The potential $V(z)$ is increasing function for all values of $C$ and $\alpha$ at value of
constant of integration $C_{1}=-1$. We plot a graph for particular values of these parameters
just for example. After present epoch  $z=0$, it increasing exponentially.

\begin{figure}[H]
\centering
\includegraphics[width=4in]{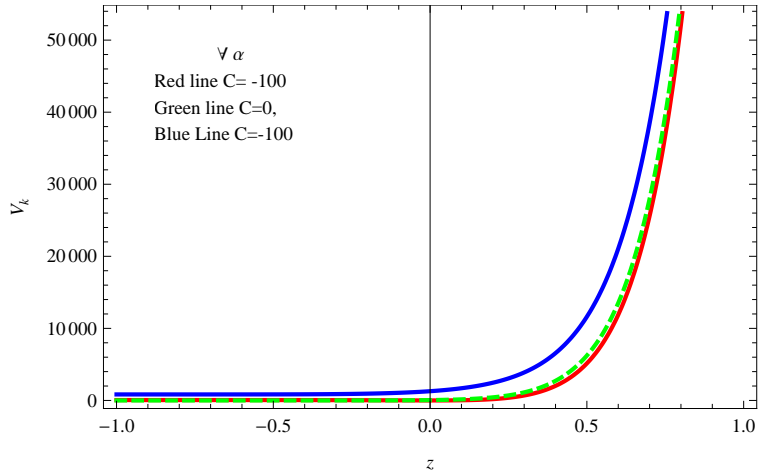}
\caption{\small{Potential vs redshift parameter}}\label{pic02}.
\end{figure}

\subsection{Dilaton Model}
The negative kinetic energy of the phantom field creates the problem of quantum instability.
To resolve this puzzle of instability, dilaton model is proposed  and further it used to study
the nature of DE. The dilaton model is defined as 4-dimensional effective low-energy model
in the context of sting theory. The pressure and energy densities are presented as
\begin{eqnarray}
\rho=-X+c_{1} e^{\lambda \phi}X^{2}, ~~~~~~~~~~ p=-X+3c_{1} e^{\lambda \phi}X^{2}.
\end{eqnarray}
where $X=\frac{\dot{\phi}^{2}}{2}$, $c_{1}$ and $\lambda$ are positive constant.
The EoS parameter $\omega=\frac{p}{\rho}$ for these densities is calculated as
\begin{eqnarray}
\omega=\frac{1-c_{1}e^{\lambda \phi}}{1-3c_{1} e^{\lambda \phi}}.
\end{eqnarray}
The comparison of Eq.(25) with Eq.(51) yields as
\begin{eqnarray}
c_{1}e^{\lambda \phi} \dot{\phi}^{2}=\frac{16C_{1}(\alpha-4)}{C^{2}t^{2}(\alpha-4)^{2}(\alpha-2)+16C_{1}(3\alpha-14)}.
\end{eqnarray}
Solving for $\phi(t)$, we reached at
\begin{eqnarray}
\phi(t)=\frac{2}{\lambda}\ln[\frac{\lambda}{2}\sqrt{\frac{16C_{1}}{C^{2}(\alpha-4)}}\sinh^{-1}(t\sqrt{\frac{C^{2}(\alpha-4)^{2}(\alpha-2)}{16C_{1}(3\alpha-14)}})].
\end{eqnarray}
In term of redshift parameter $z$
\begin{eqnarray}
\phi(z)=\frac{2}{\lambda}\log[\sqrt{\frac{16C_{1}}{C^{2}(\alpha-2)}}\sinh^{-1}
\frac{\Big(\sqrt{\frac{C^{2}(\alpha-4)^{2}(\alpha-2)}{C_{1}(3\alpha-14)}}\sqrt{\frac{4}{(1+z)^{6}}+\frac{48C_{1}}{C(4-\alpha)}}\Big)}{4\sqrt{3C(4-\alpha)}}].
\end{eqnarray}
Analytically, it is found that there is no combination of parameters $\alpha, C, C_{1}$ for which this function is defined.

\section{Summary and Discussion}
This work is devoted to study the cosmological analysis of MHRDE model in the context of
CS modified gravity. The energy density for this model is calculated and observed in
\textbf{Fig.1}. From the graph, it is obvious that density of the universe for MHRDE model is increasing for 
all values of CS modified gravity constant $C$, $\alpha<2$ and $C_{1}>0$. The deceleration parameter $q<0$
for the different combination of $C$, $C_{1}$ and $\alpha$ which advocates the accelerating expansion.
The EoS parameter $\omega<-1$ is found which favors that DE is dominant at present epoch in case
of MHRDE model in the context of CS modified gravity.

Futhermore, we reconstruct different scalar filed models using MHRDE in the context of
dynamical CS modified gravity and found interesting results plotting them graphically.
It is obvious that the potential of quintessence model depends only on the value of constant
of integration $C_{1}$. It shows the increasing behavior for all $C_{1}<0$
and decreasing for $C_{1}>0$. It is interesting that the potential in Eq.(36) is independent of
CS parameter $C$ and MHRDE parameter $\alpha$ identically, although, we are working in the
frame work of CS Modified gravity using MHRDE model. The graph plotted in \textbf{Fig.2}
for the potential of tachyon model shows the exponentially decreasing
behavior irrespective of the values of parameters $\alpha$, $C$ and $C_{1}$.
In case of k-essence, the potential is increasing function for all values
of $C$ and $\alpha$ at particular value of $C_{1}=-1$. After present epoch
$z=0$, the graph increasing exponentially given by \textbf{Fig.3}.
Analytically, it is found that there is no combination of parameters $\alpha, C, C_{1}$
for which $\phi(z)$ is defined in case of Dilaton model .
\vspace{0.5cm}

{\bf Acknowledgment}
We acknowledge the remarkable assistance of the Higher Education
Commission Islamabad, Pakistan. I would like to thanks Department of Physics and Astronomy
University of British Columbia Canada for giving me space in their Astro Lab for
six months as Visiting International Research Scholar. I really thankful to Dr.
Douglus Scott for his supervision and valuable suggestions on this work.


\begin{thebibliography}{99}
\bibitem{[1]}  A.G. Riess, et al., Supernova Search Team Collaboration, Astron. J. \textbf{116}(1998)1009;
                S. Perlmutter, et al., Supernova Cosmology Project Collaboration,Astrophys. J.\textbf{ 517}(1999)565.
\bibitem{[2]} D.N. Spergel, et al, Astrophys. J. Suppl. \textbf{170}(2007)377.
\bibitem{[3]} C.R. Contaldi, H. Hoekstra, A. Lewis, Phys. Rev. Lett. \textbf{90}(2003)221303.
\bibitem{[4]} M. Colless, et al, Mon. Not. R. Astron. Soc. \textbf{328}(2001)1039; M. Tegmark, et al, Phys. Rev. D \textbf{69}(2004)103501.
\bibitem{[5]} S. Cole, et al, Mon. Not. R. Astron. Soc. \textbf{362}(2005)505; S.P. Boughn, R.G. Crittenden, Nature \textbf{427}(2004)45.
\bibitem{[6]} S. Weinburg, Rev. Mod. Phys. \textbf{61}(1989)1;  V. Sahni and A. A. Starobinsky, Int. J. Mod. Phys. D\textbf{ 9}(2000)973.
\bibitem{[7]} E. J. Copeland, M. Sami and S. Tsujikawa, Mod. Phys. D \textbf{15}(2006)1753.
\bibitem{[8]} I. Zlatev, L.M. Wang, and P.J. Steinhardt, Phys. Rev. Lett. \textbf{82}(1999)896; X. Zhang, Mod. Phys. Lett. A \textbf{20}(2005)2575.
\bibitem{[9]} R.R. Caldwell, Phys. Lett, B \textbf{23}(2002)545; R.R. Caldwell, M. Kamionkowski and N.N. Weinberg, Phys. Rev. Lett. \textbf{91}(2003)071301.
\bibitem{[10]} T. Qiu, Mod. Phys. Lett. A \textbf{25}(2010)909; Y.F. Cai, E.N. Saridakis, M.R. Setare and J.Q. Xia, Phys. Rep. \textbf{493}(2010)1;
                M.R. Setare and E.N. Saridakis, Phys. Rev. D \textbf{79}(2009)043005.
\bibitem{[11]} C.H. Chou and K.W. Ng, Phys. Lett. B \textbf{594}(2004)1; A.G. Cohen, D.B. Kaplan and A.E. Nelson, Phys. Rev. Lett. \textbf{82}(1999)4971;
                 M. Li, Phys. Lett. B 1(2004)6.3; L.N. Granda and A. Oliveros. Phys. Lett. B \textbf{669}(2008)275.
\bibitem{[12]} C. Gao, F. Wu, X. Chen and Y.G. Shen, Phys. Rev. D,\textbf{ 79}(2009)043511.
\bibitem{[13]} L.N. Granda and A. Oliveros, Phys. Lett. B \textbf{671}(2009)199.
\bibitem{[14]} K. Karami and J. Fehri: Int. J. Theor. Phys. \textbf{49}(2010)1118.
\bibitem{[15]} R. Jackiw and S.Y. Pi, Phys. Rev. D \textbf{68}(2003)104012.
\bibitem{[16]} J. G. Silva and A.F. Santos, Eur. Phys. J. C \textbf{73}(2013)2500.
\bibitem{[17]} M.J. Amir and S. Ali, Int. J. Theor. Phys.\textbf{54}(2015)1362.
\bibitem{[18]} A. Jawad and A. Sohail, Astrophys Space Sci. \textbf{55}(2015)359.
\bibitem{[19]} S. Ali and M.J. Amir, Int. J. Theor. Phys. \textbf{DOI 10.1007/s10773-016-3131-7}.
\bibitem{[20]} A. Pasqua, R. da Rochab and S. Chattopadhyay, Eur. Phys. J. C \textbf{75}(2015)44.
\bibitem{[21]} Armendáriz-Picón, C., Mukhanov, V., Steinhardt, P.J, Phys. Rev. Lett. \textbf{85}(2000)4438.
\bibitem{[22]} G.\v{t} Hooft, arxiv preprint \textbf{gr-qc/9310026}.
\bibitem{[23]} S. Thomas, Phys. Rev. Lett.\textbf{ 89}(2002)081301.
\bibitem{[24]} R.G. Cai, Phys. Lett. B \textbf{657}(2007)228.
\bibitem{[25]} R. Bousso, J. High Energy Phys. \textbf{07}(1999)004.
\bibitem{[26]} S. Chen and J. Jing, Phys. Lett. B \textbf{679}(2009)144.
\bibitem{[27]} R.G. Cai, B. Hu and Y. Zhan, Commun. Theor. Phys. \textbf{51}(2009)954.
\bibitem{[28]} C. Gao, F. Wu, X. Chen, and Y.G. Shen, Phys. Rev. D \textbf{79}(2009)043511.
\bibitem{[29]} L.N. Granda and A. Oliveros, Phys. Lett. B \textbf{669}(2008)275.
\bibitem{[31]} C. Furtado, J. R. Nascimento, A. Y. Petrov and A. F. Santos, arXiv:\textbf{1005.1911} [hep-th].
\bibitem{[32]} Sahni, V., et al, JETP Lett. \textbf{77}(2003)201.
\bibitem{[33]} Copeland, E.J., Sami, M., Tsujikawa, S, Int. J. Mod. Phys. D \textbf{15}(2006)1753.
\bibitem{[34]} A.Sen, JHEP \textbf{0204}(2002)048.
\bibitem{[35]} A.Sen, JHEP \textbf{0207}(2002)065.
\bibitem{[36]} A.Sen, Mod. Phys. Lett. A \textbf{17}(2002)1797.
\bibitem{[37]} G.W. Gibbons, Phys. Lett. B\textbf{ 537}(2002)1.
\bibitem{[38]} S. Mukohyama, Phys. Rev. D \textbf{66}(2002)024009.
\bibitem{[39]} A. Feinstein. Phys. Rev. D \textbf{66}(2002)063511.
\bibitem{[40]} T. Padmanabhan, Phys. Rev. D \textbf{66}(2002)021301.
\bibitem{[41]} A.A. Starobinsky, JETP Lett.\textbf{68}(1998)757.




\end{thebibliography}
\end{document}